\documentclass[11pt,ams,preprintnumbers]{article}
\usepackage{geometry}
\usepackage{amsmath}
\usepackage{amsfonts}
\usepackage{amssymb}
\usepackage{graphicx}

\begin{document}

\date{}
\title{\textbf{Renormalizability of a quark-gluon model with soft BRST breaking in the infrared region}}
\author{\textbf{\ L.~Baulieu}$^{a,b}$\thanks{baulieu@lpthe.jussieu.fr}, \textbf{M.~A.~L.~Capri$^{c}$\thanks{marcio@dft.if.uerj.br}}\thinspace,
\textbf{A.~J.~G\'{o}mez$^{c}$\thanks{ajgomez@uerj.br}}\thinspace,
\textbf{V.~E.~R.~Lemes$^{c}$\thanks{vitor@dft.if.uerj.br}}\thinspace,\\
\textbf{R.~F.~Sobreiro}$^{d}$\thanks{sobreiro@if.uff.br}\;,
\textbf{S.~P.~Sorella}$^{c}$\thanks{sorella@uerj.br}\;\footnote{Work
supported by FAPERJ, Funda{\c c}{\~a}o de Amparo {\`a} Pesquisa do
Estado do Rio de Janeiro, under the program {\it Cientista do
Nosso Estado},
E-26/100.615/2007.}\\\\
$^{a}$\textit{\small Theoretical division CERN}$
$\\$^{b}$\textit{\small LPTHE, CNRS and Universit\'{e} Pierre et Marie Curie}\\
\textit{$^{c}${\small {UERJ $-$ Universidade do Estado do Rio de Janeiro}}}\\\textit{{\small {Instituto de F\'{\i}sica $-$ Departamento de F\'{\i}sica
Te\'{o}rica}}}\\\textit{{\small {Rua S{\~a}o Francisco Xavier 524, 20550-013 Maracan{\~a}, Rio
de Janeiro, Brasil}}}$$\\
\textit{$^{d}${\small {UFF $-$ Universidade Federal Fluminense, Instituto de F\'{\i}sica}}}\\\textit{{\small {Avenida Litor\^anea s/n, 24210-346 Boa Viagem, Niter\'oi, Rio de Janeiro, Brasil}}}$$}
\maketitle

\begin{flushright}
\vskip -12cm
{\small CERN-PH-TH/2009-016}\\
\vskip 11cm
\end{flushright}

\begin{abstract}
\noindent We prove the renormalizability of a quark-gluon model with a soft breaking of the BRST symmetry, which accounts for the modification of the large distance behavior of the quark and gluon correlation functions. The proof is valid to all orders of perturbation theory, by making use of softly broken Ward identities.
\end{abstract}

\baselineskip=15pt
\newpage

\section{Introduction}\label{intro}

The Gribov-Zwanziger framework
\cite{Gribov:1977wm,Zwanziger:1989mf,Zwanziger:1992qr} consists in
restricting  the domain of integration in the Feynman path
integral within  the Gribov horizon. It has motivated extended
studies of nonperturbative tools for  investigating the infrared
behavior of the gluon and ghost correlation functions, see
\cite{Alkofer:2000wg} for the use of modified Schwinger-Dyson
equations for QCD,
\cite{Dudal:2005na,Dudal:2007cw,Dudal:2008sp,Aguilar:2008xm} for
recent analytical results and
\cite{Cucchieri:2007rg,Cucchieri:2008fc,Bornyakov:2008yx} for
numerical data obtained through lattice simulations.\\\\ Zwanziger
has been able to show that the restriction of the path integral
within the Gribov horizon  for the gluon can be achieved by adding
to the Fadeev--Popov  action in the Landau gauge   a   local
action, depending on new fields, with well-defined interactions
with the gluons  and their Faddeev-Popov  ghosts in the Landau
gauge \cite{ Zwanziger:1989mf,Zwanziger:1992qr}. However, this
local action violates the BRST symmetry by a soft term, so that
BRST symmetry is only enforced in the scaling limit.  The yet
unorthodox point of view that the BRST symmetry can be broken in
the IR region of QCD  was heuristically  anticipated  by Fujikawa
\cite{Fujikawa:1982ss}. It is not in contradiction with any given
physical principle, since it is by no means necessary that the QCD
microscopic theory possesses a unitary sector for its partons,
namely the  quarks and gluons,  to warrantee unitarity properties
among the sector of bound states that constitute its spectrum. The
only requirement is that the modified theory remains
renormalizable  and that the BRST symmetry is recovered in the
ultraviolet region, in order to suitably describe the asymptotic
properties in terms of almost deconfined partons, as predicted by
asymptotic freedom and short distance expansion. From a physical
point of view, it is gratifying for the quark-gluon model
introduced in \cite{Baulieu:2008fy} that the modification of the
usual Feynman propagators into a Gribov-type propagators
eliminates from the beginning all partons from  the spectrum,
since their modified propagators have no poles on the real axis, a
property  that anticipates  quite well the confinement. \\\\What
justified our previous work \cite{Baulieu:2008fy}  is  that the
genuine geometrical approach of Zwanziger leaves aside the quarks,
which do not participate to the Gribov phenomenon, while the idea
of a parton model suggests that  quarks and gluons should be
treated on the same footing, and quark propagators should have  an
analogous behavior as the gluon one in the infra-red domain. An
idea for getting such modified quark propagators was thus needed,
which goes beyond the Gribov question.  A generalization of the
work of Zwanziger was also needed
\cite{Dudal:2007cw,Dudal:2008sp,Aguilar:2008xm} in order to
achieve a different behavior for the shape of the genuine
Gribov-gluon propagator $D(q^2)$ that vanishes at $q^2=0$, for
accommodating recent lattice simulations that seem to indicate
that the gluon propagator goes to a non-vanishing constant at very
small $q^2$
\cite{Cucchieri:2007rg,Cucchieri:2008fc,Bornyakov:2008yx}.

\noindent To provide such an improved, local and renormalizable,
quantum field theory that gives the wanted modifications in the
infra-red region, both for quarks and gluons, we proposed in
\cite{Baulieu:2008fy} the following picture.  Given a  theory of
partons (eg   quark and gluons in four  dimensions, or a scalar
fields in dimensions such that the ultraviolet divergences are
renormalizable or super-renormalizable),  it can always be coupled
to a topological field theory made of new fields arranged as BRST
trivial doublets, in such a way that the partons are already
confined at the tree level  by their mixing with the unphysical
fields of the topological field theory. We found that such a
mixing can be generally  allowed by a soft breaking of the BRST
symmetry. Here, confinement is meant in a very simple way. The
propagators  of all fields have only poles at complex
positions\footnote{More precisely, the propagators display
violation of reflection positivity, a feature which invalidates
the interpretation of partons as excitations of the physical
spectrum of the theory.}. This implies that the theory has no
vacuum for the partons and all observables are made of composite
operators, defined by solving the cohomology of the BRST operator.
These composite operators can be renormalized in the standard way,
with expectation values related to the parameters of the soft
breaking mechanism. This idea was inspired by the algebraic
characterization of the local terms that Zwanziger introduced to
complete the Faddeev--Popov action for the gluons. Part of the
Zwanziger action can be recognized as a topological action,  which
involves bosons and fermions that transform under the BRST
symmetry as  a system of two BRST trivial doublets, and it is
BRST-exact. The remaining part of Zwanziger's action breaks the
BRST symmetry in a soft way, and yields a Gribov propagator for
the transverse gluon.
\\\\The addition of the new fields arranged as BRST-exact doublets
eventually allows for the introduction of massive parameters that
can be used to modify the infrared behavior of the theory, without
changing the set of observables. The necessity of breaking the
BRST symmetry can be easily understood within this framework.  If
the added action were BRST-exact, nothing would be changed for the
predictions of  the original parton theory. Indeed, the
observables, defined as the cohomology of the BRST operator, are
the same and the effect of integrating over the new fields  would
only consist in multiplying the partition function by one. So, if
the new field dependence  is through BRST-invariant terms (and
thus BRST-exact terms because they transform as BRST-exact
doublets), there is no way to improve the infra-red behavior of
the amplitudes. One can notice that the possibility  of adding
such a BRST-exact terms can be related to the  formal  invariance
of the path integral under arbitrary changes of variables for  the
parton fields. By doing such changes of variables and  playing
with Lagrange multipliers and determinants formula, one can indeed
recover the class of BRST-exact action that are bilinear in the
new fields that have the opposite statistics to the
partons\footnote{In string theory, this idea has been already used
by adding so-called "topological packages" to any given 2d-string
action, which allows one to show in this way possible
relationships between the different string models
\cite{Baulieu:1996hn,Baulieu:1996mr,Baulieu:1996ur}.}. To obtain the wanted modification of the
infra-red behavior of parton correlators,  one  must go further,
and consider the possibility of an explicit soft breaking of the
BRST symmetry. This, of course, can  modify  the parton
propagators in the low energy domain, without changing their
ultraviolet behavior, provided the theory is renormalizable. This
will also modify the numerical values of the observables, by
giving them a dependence from the parameters of the soft breaking.
\\\\In the case of the improved quark-gluon theory, the properties
of the breaking and the respective consequences for the theory can
be summarized as follows:
\begin{itemize}
\item {The breaking term is  soft, meaning that its dimension is
smaller than the space-time dimension. As a consequence, the
breaking can be neglected in the ultraviolet region, where one
recovers the notion of exact BRST symmetry.}

\item {The BRST operator preserves nilpotency. Moreover, the set
of physical operators of the theory, identified with the
cohomology classes of the BRST operator, is left unmodified. This
occurs because the new additional fields are introduced as BRST
doublets. As such, they do not alter the cohomology of the BRST
operator.}

\item {The soft breaking of the BRST symmetry is introduced in a
way compatible with the renormalizability. More precisely, it is
associated to a quadratic term in the fields obtained by demanding
that the additional fields couple linearly to the original fields
of the theory, so that the resulting propagators turn out to be
modified in the infrared region. As a consequence, a given
correlation function can display a different behavior, when going
from the deep ultraviolet to the infrared region in momentum
space.}

\item {The soft BRST symmetry breaking is meant to be an explicit
breaking, {\it i.e.} it is not a spontaneous symmetry breaking,
which would give rise to Goldstone massless fields, and thus to a
quite different framework, as it will be explained elsewhere.}

\item All massive parameters should be related to the unique scale
of the theory, namely $\Lambda_{\rm QCD}$, by requiring that the
massive soft parameters satisfy suitably gap type equations which
allow to determine them in a self-consistent way. This is the
case, for example, of the massive Gribov parameter $\gamma$ which
is fixed by a gap equation
\cite{Zwanziger:1989mf,Zwanziger:1992qr,Dudal:2005na,Dudal:2007cw,Dudal:2008sp},
see Sect.4.

\end{itemize}

\noindent Following the procedure described above, in
\cite{Baulieu:2008fy} a model accounting for a soft BRST symmetry
breaking giving rise to a modification of the long distance
behavior of the quark propagator was established. In fact, the
model predicts
\begin{equation}
\left\langle\psi(k){\bar
\psi}(-k)\right\rangle=\frac{i\gamma_{\mu}k_{\mu}+{\cal
A}(k)}{k^{2}+{\cal A}^{2}(k)}\;,\label{d}
\end{equation}
instead of the standard Dirac propagator for the quarks
$\frac{i\gamma_{\mu}k_{\mu}}{k^2}$. In this equation
\begin{equation}
{\cal A}(k)=\frac{2M_{1}^{2}M_{2}}{k^{2}+m^{2}}\;,\label{e}
\end{equation}
is a function depending on the soft breaking mass parameters
$M_{1}$, $M_{2}$ and $m$. From expression \eqref{e} one sees that
the function ${\cal A}(k)$ vanishes in the deep ultraviolet region
where the usual perturbative behavior for the propagator is
recovered. It is worth mentioning that expression \eqref{e}
provides a good fit for the dynamical mass generation for quarks
in the infrared region in the Landau gauge, as reported by lattice
numerical simulations of the quark propagator
\cite{Parappilly:2005ei,Furui:2006ks}. We recall that the function
\eqref{e} is analogous to the one appearing in the gluon
propagator within the Gribov-Zwanziger framework
\cite{Zwanziger:1989mf,Zwanziger:1992qr}. In fact, as discussed in
\cite{Baulieu:2008fy}, the Gribov-Zwanziger action can be
recovered through the introduction of a soft breaking, related to
the appearance of the Gribov parameter $\gamma$. This parameter is
needed to implement the restriction to the Gribov horizon, which
turns out to be at the origin of the soft breaking of the BRST
symmetry \cite{Dudal:2007cw,Dudal:2008sp}. In particular,   for
the tree level gluon propagator, one finds
\begin{equation}
\left\langle
A_\mu^a(k)A_\nu^b(-k)\right\rangle=\delta^{ab}\left(\delta_{\mu\nu}
-\frac{k_\mu
k_\nu}{k^2}\right)\frac{1}{k^{2}+{\cal{M}}^{2}(k)}\;,\label{d00}
\end{equation}
with
\begin{equation}
{\cal M}^2(k)=\frac{\gamma^4}{k^2+\mu^2}\;,\label{e00}
\end{equation}
where the second mass parameter $\mu$ accounts for the nontrivial
dynamics of the auxiliary fields needed to localize Zwanziger's
horizon function, see \cite{Dudal:2007cw,Dudal:2008sp}. \\\\It is
of course striking how similar are the modified gluon and quark
propagators, in the sense that the introduction of the massive
parameters by the above mechanism has eliminated the infra-red
problem for both type  of fields. Within the context of plain
perturbation theory, these parameters can  be fined-tuned, and
even be used as plain infra-red regulators that are controlled by
softly broken Ward identities. However, in a confining theory like
QCD, they must be related to physical observables, and
non-perturbatively computed as a function  of  the basic parameter
of the theory, $\Lambda_{\rm QCD}$.  The clear advantage of our
approach is that  the  modified theory is no more submitted to the
Gribov ambiguity and   the parton (quark and gluons)  propagators
have no poles at real values.  The  model excludes their
appearance in the spectrum from the beginning, and its
perturbative expansion can be possibly compared to the prediction
of non-perturbative approach such as the lattice formulation in
the Landau gauge.  The method solves in a quite  simple conceptual
way the necessity of introducing  massive parameters, not only for
the Gribov ambiguities but also for defining the scale of the
chiral symmetry breaking. \\\\The goal of this paper is to prove
an elementary, but necessary property of this model:  it has to be
multiplicatively renormalizable, to ensure that the breaking of
the BRST symmetry remains  soft, at any given finite order of
perturbation theory, and that no further parameters than those
which modify the infra-red behavior of propagators can appear. We
will do this by employing the algebraic renormalization
\cite{Piguet:1995er}. Notice that such  a proof can be generalized
to other cases, where one may wish to introduce  an infra-red
cut-off for  other  theories with nontrivial  gauge invariance and
control its effect by softly broken Ward identities, within the
context of locality. This could be of relevance  for the study of
non-exactly  solvable super-renormalizable theories, like
3-dimensional  gauge theories, where one must control the way the
infra-red regularization is compatible with the BRST symmetry, or
even for gravity, when it is computed at a finite order of
perturbation theory. \\\\The paper is organized as follows. In
Section \ref{the_model} we provide a short overview of the model
and of its soft BRST symmetry breaking. In Section \ref{proof} the
algebraic proof of the renormalizability is given. In Section
\ref{GZ_term} we consider the inclusion of the Gribov-Zwanziger
action and we discuss the renormalizability of the resulting
model. Finally, in Section \ref{conclusion} we collect our
conclusions.

\section{The quark model and its BRST soft breaking term}\label{the_model}

We start with the Yang-Mills action quantized in the Landau gauge,
\begin{equation}
S_{\mathrm{inv}}=\int d^{4}x\left(  \frac{1}{4}F_{\mu\nu}^{a}F_{\mu\nu}
^{a}+\bar{\psi}_{\alpha}^{i}\left(  \gamma_{\mu}\right)  _{\alpha\beta}D_{\mu
}^{ij}\psi_{\beta}^{j}+ib^{a}\partial_{\mu}A_{\mu}^{a}+\bar{c}^{a}
\partial_{\mu}D_{\mu}^{ab}c^{b}\right)\;,\label{1}
\end{equation}
where
\begin{equation}
D_{\mu}^{ij}=\partial_{\mu}\delta^{ij}-ig(T^{a})^{ij}A_{\mu}^{a}\;,\label{dc0}
\end{equation}
is the covariant derivative in the fundamental representation of the $SU(N)$
gauge group, with generators $(T^{a})^{ij}$, and
\begin{equation}
D_{\mu}^{ab}=\partial_{\mu}\delta^{ab}-gf^{abc}A_\mu^c\;,\label{dc1}
\end{equation}
is the covariant derivative in the adjoint representation. The
first set of small Latin indices $\{i,j,\ldots\}\in\{1,..,N\}$
will be used to denote the fundamental representation, while the
second set $\{a,b,\ldots,h\}\in\{1,..,N^{2}-1\}$ will be employed
for the adjoint representation. The set of Greek indices
$\{\alpha,\beta,\gamma,\delta\}$ stand for spinor indices. The
remaining Greek indices will denote space-time indices.\\\\
The action \eqref{1} is invariant under the BRST transformations, namely
\begin{eqnarray}
sA_\mu^a&=&-D_\mu^{ab}c^b\;,\nonumber\\
s\psi_\alpha^i&=&-igc^a\left(T^a\right)^{ij}\psi_\alpha^j\;,\nonumber\\
s\bar{\psi}_\alpha^i&=&-ig\bar{\psi}_\alpha^jc^a(T^a)^{ji}\;,\nonumber\\
sc^a&=&\frac{1}{2}gf^{abc}c^bc^c\;,\nonumber\\
s\bar{c}^a&=&ib^a\;,\nonumber\\
sb^a&=&0\;.\label{2}
\end{eqnarray}\\\\
Following \cite{Baulieu:2008fy}, a model for the dynamical quark
mass generation can be constructed by introducing two BRST
doublets of spinor fields $(\xi^{i},\theta^{i})$ and
$(\eta^{i},\lambda^{i})$, transforming as
\begin{eqnarray}
s\xi_\alpha^i&=&\theta_\alpha^i\;,\;\;s\theta_\alpha^i\;\;=\;\;0\;,\nonumber\\
s\eta_\alpha^i&=&\lambda_\alpha^i\;,\;\;s\lambda_\alpha^i\;\;=\;\;0\;.\label{z}
\end{eqnarray}
The propagation of these fields is described by the following BRST-exact action
\begin{eqnarray}
S_{\xi\lambda}&=&s\int d^4x\left(-\bar{\eta}_\alpha^i\partial^2\xi_\alpha^i+
\bar{\xi}_\alpha^i\partial^2\eta_\alpha^i+m^2(\bar{\eta}_\alpha^i\xi_\alpha^i
-\bar{\xi}_\alpha^i\eta_\alpha^i)\right)\nonumber\\
&=&\int d^4x\left(-\bar{\lambda}_\alpha^i\partial^2\xi_{\alpha}^i-\bar{\xi}_\alpha^i\partial^2\lambda_\alpha^i-
\bar{\eta}_\alpha^i\partial^2\theta_{\alpha}^i+\bar{\theta}_\alpha^i\partial^2\eta_\alpha^i+
m^2(\bar{\lambda}_\alpha^i\xi_{\alpha}^i+\bar{\xi}_\alpha^i\lambda_\alpha^i+\bar{\eta}_\alpha^i\theta_\alpha^i-
\bar{\theta}_\alpha^i\eta_\alpha^i)\right)\;,\label{G}
\end{eqnarray}
where $m$ is a mass parameter. Further, we introduce the coupling of the spinors $(\xi^i,\theta^i)$ and $(\eta^i,\lambda^i)$ with the matter field $\psi_\alpha^i$,
\begin{equation}
S_{M}=\int d^4x\left(M_1^2(\bar{\xi}_\alpha^i\psi_\alpha^i+\bar{\psi}_\alpha^i\xi_\alpha^i)-
M_2(\bar{\lambda}_\alpha^i\psi_\alpha^i+\bar{\psi}_\alpha^i\lambda_\alpha^i)\right)\;.\label{g}
\end{equation}
Evidently, the action \eqref{G} is BRST-invariant. This is not the
case of $S_{M}$, which will give rise to a soft breaking of the
BRST symmetry, parameterized by the two soft mass parameters
$M_1$, $M_2$. In fact,
\begin{equation}
sS_M=\Delta\;,\label{delta}
\end{equation}
where
\begin{eqnarray}
\Delta&=&\int d^{4}x
\left(M_1^2(\bar{\theta}_\alpha^i\psi_\alpha^i-\bar{\psi}_\alpha^i
\theta_\alpha^i)+igM_1^2c^a\left(\bar{\xi}^i_\alpha(T^a)^{ij}\psi^j_\alpha-\bar{\psi}^i_\alpha(T^a)^{ij}\xi^j_\alpha\right)\right.\nonumber\\
&{\ }& {\ }{\ }{\ }{\ }
+\left.igM_{2}c^a(\bar{\lambda}_\alpha^i(T^a)^{ij}\psi_{\alpha}^{j}-\bar{\psi}_\alpha^i(T^a)^{ij}\lambda_\alpha^j)\right)\;.\label{bk}
\end{eqnarray}
Notice that $\Delta$ is a soft breaking, \emph{i.e.} it is of dimension less than four.\\\\
Performing the integration over the auxiliary spinor fields,
yields a nonlocal action for $\psi$ and $\bar{\psi}$,
\begin{equation}
S_{\psi}=\int d^{4}x\left( \bar{\psi}_{\alpha}^{i}\left(  \gamma_{\mu}\right)  _{\alpha\beta}D_{\mu
}^{ij}\psi_{\beta}^{j} -2M_{1}^{2} M_{2}\bar{\psi}_{\alpha}^{i} \left( \frac{1}{\partial^{2}-m^{2}}\right) \psi_{\alpha}^{i}\right)\;.\label{integrated0}
\end{equation}
It is important to emphasize that the introduction of the
auxiliary spinor fields can be seen as a tool for localizing the
nonlocal term \eqref{integrated0} that appears in the fermionic
sector, in the same way as the nonlocal Gribov-Zwanziger horizon
function is cast in local form through the introduction of a
suitable set of auxiliary BRST doublet fields
\cite{Zwanziger:1989mf,Zwanziger:1992qr,Dudal:2005na,Dudal:2008sp}.

\subsection{Introducing sources for controlling the soft symmetry breaking}

In order to prove that the model described by the action
\begin{equation}
S=S_{\mathrm{inv}}+S_{\xi\lambda}+S_{M}\;,\label{s1}
\end{equation}
is multiplicatively renormalizable, we follow the procedure
outlined by Zwanziger in the study of the Gribov horizon in the
Landau gauge \cite{Zwanziger:1989mf,Zwanziger:1992qr}. It amounts
to embedding the action \eqref{s1} in a larger model,
$S\rightarrow S_0$, displaying exact BRST invariance. This is
achieved by treating the breaking term $\Delta$ as a composite
operator, which is introduced into the theory through a suitable
set of external sources. The original action $S$ is thus recovered
from the extended action $S_0$ by demanding that the sources
acquire a particular value, which we shall refer to as the
physical value. The renormalizability of $S$ follows thus by
proving the renormalizability
of the extended action $S_0$.\\\\
In order to introduce the extended invariant action $S_0$, we make
use of the following set of sources $(J,H)$, $(\bar{J},\bar{H})$,
$(K,G)$, $(\bar{K},\bar{G})$ and $\left(N,P\right)$, assembled in
BRST doublets\footnote{For future purpose, we recall that all
quantum numbers of fields and sources are displayed in Tables
\ref{table1} and \ref{table2}, including the  charge $Q_{4N}$ that
will be defined through expression \eqref{qop}.}, {\it i.e.}
\begin{eqnarray}
sJ_{\alpha\beta}^{ij}&=&H_{\alpha\beta}^{ij}\;,\;\;sH_{\alpha\beta}^{ij}\;\;=\;\;0,\nonumber\\
s\bar{J}_{\alpha\beta}^{ij}&=&\bar{H}_{\alpha\beta}^{ij}\;,\;\;s\bar{H}_{\alpha\beta}^{ij}\;\;=\;\;0,\nonumber\\
sK_{\alpha\beta}^{ij}&=&G_{\alpha\beta}^{ij}\;,\;\;sG_{\alpha\beta}^{ij}\;\;=\;\;0,\nonumber\\
s\bar{K}_{\alpha\beta}^{ij}&=&\bar{G}_{\alpha\beta}^{ij}\;,\;\;s\bar{G}_{\alpha\beta}^{ij}\;\;=\;\;0,\nonumber\\
sN&=&P\;,\;\;\;\;\;\;\;\;\;sP\;\;=\;\;0.\label{4}
\end{eqnarray}
The invariant action that accounts for the extra spinor fields and
for the breaking term is thus defined as
\begin{eqnarray}
S_{JK}&=&s\int d^{4}x\left(-\bar{\eta}_\alpha^i\partial^2
\xi_\alpha^i+\bar{\xi}_\alpha^i\partial^2\eta_\alpha^i+P(\bar{\eta}_\alpha^i\xi_\alpha^i-
\bar{\xi}_\alpha^i\eta_\alpha^i)+\sigma NP\right.\nonumber\\
&+&\left.J_{\alpha\beta}^{ij}\bar{\xi}_\alpha^i\psi_\beta^j+\bar{J}_{\alpha\beta}^{ij}\bar{\psi}_\beta^j
\xi_\alpha^i+K_{\alpha\beta}^{ij}\bar{\lambda}_\alpha^i\psi_\beta^j+\bar{K}_{\alpha\beta}^{ij}\bar{\psi}_\beta^j\lambda_\alpha^i\right)\;,\label{fa}
\end{eqnarray}
which, explicitly, reads
\begin{eqnarray}
S_{JK}&=&\int
d^{4}x\left(-\bar{\lambda}_\alpha^i\partial^2\xi_\alpha^i-\bar{\xi}_\alpha^i\partial^2\lambda_\alpha^i-
\bar{\eta}_\alpha^i\partial^2\theta_\alpha^i+\bar{\theta}_\alpha^i\partial^2\eta_\alpha^i+P(\bar{\lambda}_\alpha^i\xi_\alpha^i+
\bar{\xi}_\alpha^i\lambda_\alpha^i+\bar{\eta}_\alpha^i\theta_\alpha^i-\bar{\theta}_\alpha^i\eta_\alpha^i)+\sigma P^2\right.\nonumber\\
&+&\left.H_{\alpha\beta}^{ij}\bar{\xi}_\alpha^i\psi_\beta^j+\bar{H}_{\alpha\beta}^{ij}\bar{\psi}_\beta^j
\xi_\alpha^i-J_{\alpha\beta}^{ij}\left(\bar{\theta}_\alpha^i\psi_\beta^j-\bar{\xi}_\alpha^iigc^a\left(T^a\right)^{jk}\psi_\beta^k\right)-
\bar{J}_{\alpha\beta}^{ij}\left(ig\bar{\psi}_\beta^kc^a\left(T^a\right)^{kj}\xi_\alpha^i-\bar{\psi}_\beta^j\theta_\alpha^i\right)\right.
\nonumber\\
&+&\left.G_{\alpha\beta}^{ij}\bar{\lambda}_\alpha^i\psi_\beta^j+\bar{G}_{\alpha\beta}^{ij}\bar{\psi}_\beta^j\lambda_\alpha^i+
K_{\alpha\beta}^{ij}\bar{\lambda}_\alpha^iigc^a\left(T^a\right)^{jk}\psi_\beta^k-\bar{K}_{\alpha\beta}^{ij}ig
\bar{\psi}_\beta^kc^a\left(T^a\right)^{kj}\lambda_\alpha^i\right)\;,\label{f}
\end{eqnarray}
whose BRST invariance is manifest. The parameter $\sigma$, in
expression \eqref{f}, is a dimensionless parameter needed for
renormalization purposes.  For the so called physical values of
the external sources, we have
\begin{eqnarray}
J_{\alpha\beta}^{ij}\Big|_{\mathrm{phys}}&=&\bar{J}_{\alpha\beta}^{ij}\Big|_{\mathrm{phys}}\;\;=\;\;0\;,\nonumber\\
K_{\alpha\beta}^{ij}\Big|_{\mathrm{phys}}&=&\bar{K}_{\alpha\beta}^{ij}\Big|_{\mathrm{phys}}\;\;=\;\;0\;,\nonumber\\
H_{\alpha\beta}^{ij}\Big|_{\mathrm{phys}}&=&\bar{H}_{\alpha\beta}^{ij}\Big|_{\mathrm{phys}}\;\;=\;\;M_1^2\delta^{ij}\delta_{\alpha\beta}\;,\nonumber\\
G_{\alpha\beta}^{ij}\Big|_{\mathrm{phys}}&=&\bar{G}_{\alpha\beta}^{ij}\Big|_{\mathrm{phys}}\;\;=\;\;-M_2\delta^{ij}
\delta_{\alpha\beta}\;,\nonumber\\
P\Big|_{\mathrm{phys}}&=&m^{2}\;,\nonumber\\
N\Big|_{\mathrm{phys}}&=&0\;.\label{A}
\end{eqnarray}\\\\
The BRST-invariant extended action $S_0$ is thus defined as
\begin{equation}
S_0=S_{\mathrm{inv}}+S_{\xi\lambda}+S_{JK}+S_{ext}\;.\label{5}
\end{equation}
It is easily checked that the starting action $S$, eq.\eqref{s1},
is recovered from the extended action $S_0$ when taking the
physical values of the sources, eq.\eqref{A}, namely
\begin{equation}
S_0\Big|_{\mathrm{phys}}=S+\int{d^4x}\;\sigma m^4\;.\label{pl}
\end{equation}

\begin{table}[t]
\centering
\begin{tabular}{|c|c|c|c|c|c|c|c|c|c|c|c|c|c|c|}
\hline
fields & $A$ & $b$ & $c$ & $\bar{c}$ & $\psi$ & $\bar{\psi}$ & $\xi$ & $\bar{\xi}$ & $\lambda$ & $\bar{\lambda}$ & $\theta$ & $\bar{\theta}$ & $\eta$ & $\bar{\eta}$ \\
\hline
UV dimension & 1 & 2 & 0 & 2 & $3/2$ & $3/2$ & $1/2$ & $1/2$ & $3/2$ & $3/2$ & $1/2$ & $1/2$ & $3/2$ & $3/2$\\
Ghost number & 0 & 0 & 1 & $-1$ & 0 & 0 & 0 & 0 & 0 & 0 & 1 & 1 & $-1$ & $-1$\\
$Q_{4N}$ - charge & 0 & 0 & 0 & 0 & 0 & 0 & 1 & $-1$ & 1 & $-1$ & 1 & $-1$ & 1 & $-1$\\
Spinor number & 0 & 0 & 0 & 0 & 1 & $-1$ & 0 & 0 & 0 & 0 & 0 & 0 & 0 & 0\\
Statistics & 0 & 0 & 1 & $-1$ & 1 & $-1$ & 1 & $-1$ & 1 & $-1$ & 2 & 0 & 0 & $-2$\\
\hline
\end{tabular}
\caption{Quantum numbers of the fields.}
\label{table1}
\end{table}

\begin{table}[t]
\centering
\begin{tabular}{|c|c|c|c|c|c|c|c|c|c|c|c|c|c|c|}
\hline
sources & $\Omega$ & $L$ & $Y$ & $\bar{Y}$ & $J$ & $\bar{J}$ & $K$ & $\bar{K}$ & $H$ & $\bar{H}$ & $G$ & $\bar{G}$ & $N$ & $P$ \\
\hline
UV dimension & 3 & 4 & $5/2$ & $5/2$ & 2 & 2 & 1 & 1 & 2 & 2 & 1 & 1 & 2 & 2 \\
Ghost number & $-1$ & $-2$ & $-1$ & $-1$ & $-1$ & $-1$ & $-1$ & $-1$ & 0 & 0 & 0 & 0 & $-1$ & 0\\
$Q_{4N}$ - charge & 0 & 0 & 0 & 0 & 1 & $-1$ & 1 & $-1$ & 1 & $-1$ & 1 & $-1$ & 0 & 0\\
Spinor number & 0 & 0 & 1 & $-1$ & 1 & $-1$ & 1 & $-1$ & 1 & $-1$ & 1 & $-1$ & 0 & 0\\
Statistics & $-1$ & $-2$ & 0 & $-2$ & 1 & $-3$ & 1 & $-3$ & 2 & $-2$ & 2 & $-2$ & $-1$ & 0\\
\hline
\end{tabular}
\caption{Quantum numbers of the sources.}
\label{table2}
\end{table}

\noindent It is worth noticing that the action $S_0$ possesses an
additional $U(4N)$ global symmetry, provided by
\begin{equation}
Q_{\alpha\beta}^{ij}S_0=0\;,\label{q00}
\end{equation}
where
\begin{eqnarray}
Q_{\alpha\beta}^{ij}&=&\int d^{4}x\left(\xi_{\alpha}^{i}%
\frac{\delta}{\delta\xi_{\beta}^{j}}-\bar{\xi}_{\alpha}^{i}%
\frac{\delta}{\delta\bar{\xi}_{\beta}^{j}}+\lambda_{\beta}^{j}\frac
{\delta}{\delta\lambda_{\alpha}^{i}}-\bar{\lambda}_{\beta}^{j}\frac{\delta
}{\delta\bar{\lambda}_{\alpha}^{i}}+\theta_{\alpha}^{i}\frac{\delta}%
{\delta\theta_{\beta}^{j}}-\bar{\theta}_{\alpha}^{i}\frac{\delta}{\delta
\bar{\theta}_{\beta}^{j}}+\eta_{\beta}^{j}\frac{\delta}{\delta\eta_{\alpha
}^{i}}-\bar{\eta}_{\beta}^{j}\frac{\delta}{\delta\bar{\eta}_{\alpha}^{i}%
}+J_{\beta\sigma}^{jk}\frac{\delta}{\delta J_{\alpha\sigma}^{ik}}\right.
\nonumber\\
&-&\left.\!\!\bar{J}_{\beta\sigma}^{jk}\frac{\delta}{\delta\bar{J}%
_{\alpha\sigma}^{ik}}+K_{\alpha\sigma}^{ik}\frac{\delta}{\delta K_{\beta
\sigma}^{jk}}-\bar{K}_{\alpha\sigma}^{ik}\frac{\delta}{\delta\bar{K}%
_{\beta\sigma}^{jk}}+G_{\alpha\sigma}^{ik}\frac{\delta}{\delta G_{\beta\sigma
}^{jk}}-\bar{G}_{\alpha\sigma}^{ik}\frac{\delta}{\delta\bar{G}_{\beta\sigma
}^{jk}}+H_{\beta\sigma}^{jk}\frac{\delta}{\delta H_{\alpha\sigma}^{ik}}%
-\bar{H}_{\alpha\sigma}^{ik}\frac{\delta}{\delta\bar{H}_{\beta\sigma}^{jk}%
}\right).\label{qop}%
\end{eqnarray}
The trace of the operator $Q_{\alpha\beta}^{ij}$, {\it i.e.}
$Q_{ii}^{\alpha\alpha}=Q_{4N}$, defines a new conserved charge,
and thus an additional quantum number for the fields and sources,
allowing for the introduction of a multi-index $I=(i,\alpha)$,
with $I\in\{1,\ldots,4N\}$, for the fields
$\phi=\left(\xi,\eta,\lambda,\theta\right)$ and sources
$\Gamma=\left(K,J,H,G\right)$. Accordingly, we shall set
\begin{eqnarray}
\phi_\alpha^i&=&\phi^I\;,\nonumber\\
\Gamma_{\alpha\beta}^{ij}&=&\Gamma_\beta^{Ij}\;.
\end{eqnarray}
From now on, we shall make use of the multi-index notation.

\section{Algebraic proof of the renormalizability}\label{proof}

Let us face now the issue of the renormalizability of the extended
action $S_0$, a task which we shall undertake by making use of the
algebraic renormalization \cite{Piguet:1995er}. Let us start by
establishing the set of Ward identities fulfilled by the action
$S_0$. To that purpose, we add external sources, $L^a$,
$\Omega_\mu^a$, $Y_\alpha^i$ and $\bar{Y}_\alpha^i$, coupled to
the non-linear BRST variations of the fields, namely
\begin{eqnarray}
S_{ext}&=&s\int d^{4}x\left(\Omega_\mu^aA_\mu^a+L^ac^a+\bar{Y}_{\alpha}^{i}\psi_{\alpha}^{i}+\bar{\psi}_{\alpha}^iY_\alpha^i\right)\nonumber\\
&=&\int d^{4}x\left(-\Omega_{\mu}^aD_\mu c^a+\frac{1}{2}gf^{abc}L^ac^bc^c-ig\bar{Y}_{\alpha}^{i}c^{a}\left(  T^{a}\right)
^{ij}\psi_{\alpha}^{j}-ig\bar{\psi}_{\alpha}^{j}c^{a}\left(  T^{a}\right)
^{ji}Y_\alpha^i\right)\;,\label{m}
\end{eqnarray}
with
\begin{equation}
s\Omega_{\mu}^{a}=sL^{a}=sY=s\bar{Y}=0\;.\label{brs1}%
\end{equation}
The term \eqref{m} allows one to convert the BRST symmetry into
the corresponding Slavnov-Taylor identity. Thus, to prove the
renormalizability of the model we shall consider the more general
action
\begin{equation}
\Sigma=S_0+S_{ext}\;.\label{sigma}
\end{equation}

\subsection{Ward Identities}

The complete action \eqref{sigma} fulfills a rich set of Ward
Identities, namely:
\begin{itemize}
\item The  Slavnov-Taylor identity
\begin{equation}
{\cal S}(\Sigma)=0\;,\label{st}
\end{equation}
where
\begin{eqnarray}
{\cal S}(\Sigma)&=&\int d^{4}x\left(\frac{\delta\Sigma}{\delta\Omega_{\mu}^{a}%
}\frac{\delta\Sigma}{\delta A_{\mu}^{a}}+\frac{\delta\Sigma}{\delta\bar
{Y}_{\alpha}^{i}}\frac{\delta\Sigma}{\delta\psi_{\alpha}^{i}}-\frac
{\delta\Sigma}{\delta Y_{\alpha}^{i}}\frac{\delta\Sigma}{\delta\bar{\psi
}_{\alpha}^{i}}+\frac{\delta\Sigma}{\delta c^{a}}\frac{\delta\Sigma}{\delta
L^{a}}+ib^{a}\frac{\delta\Sigma}{\delta\bar{c}^{a}}+\bar{\theta}^{I} \frac{\delta\Sigma}%
{\delta\bar{\xi}^{I}}+\bar{\lambda}^{I}\frac{\delta\Sigma}{\delta\bar{\eta
}^{I}}\right.\nonumber\\
&+&\left.\theta^{I}\frac{\delta\Sigma}{\delta\xi^{I}}%
+\lambda^{I}\frac{\delta\Sigma}{\delta\eta^{I}}+H_{\alpha}^{Ij}\frac{\delta\Sigma}{\delta
J_{\alpha}^{Ij}}+G_{\alpha}^{Ij}\frac{\delta\Sigma}{\delta K_{\alpha}^{Ij}%
}+\bar{H}_{\alpha}^{Ij}
\frac{\delta\Sigma}{\delta\bar{J}_{\alpha}^{Ij}}
+\bar{G}_{\alpha}^{Ij}\frac{\delta\Sigma}{\delta\bar{K}_{\alpha}^{Ij}}
+P\frac{\delta\Sigma}{\delta N}\right)\;.
\end{eqnarray}

\item The gauge condition:
\begin{equation}
\frac{\delta\Sigma}{\delta b^{a}}=i\partial_{\mu}A_{\mu}^{a}\;.
\end{equation}

\item  The antighost equation:
\begin{equation}
\frac{\delta\Sigma}{\delta\bar{c}^{a}}+\partial_{\mu}\frac{\delta\Sigma}{\delta\Omega_{\mu}^{a}}=0\;.
\end{equation}

\item The ghost equation:
\begin{equation}
{\cal G}^{a}\Sigma=\Delta_{cl}^{a}\;,\label{ghost}
\end{equation}
with
\begin{eqnarray}
{\cal G}^{a}&=&\int d^{4}x\left(  \frac{\delta}{\delta
c^{a}}-if^{abc}\bar {c}^{b}\frac{\delta}{\delta b^{c}}-ig\left(
T^{a}\right) ^{jk}\left(
J_{\alpha}^{Ij}\frac{\delta}{\delta H_{\alpha}^{Ik}}+K_{\alpha}^{Ij}%
\frac{\delta}{\delta G_{\alpha}^{Ik}}\right)  \right. \nonumber \\
&-&\left.ig\left(  T^{a}\right)  ^{kj}\left(  \bar{J}_{\alpha}^{Ij}%
\frac{\delta}{\delta\bar{H}_{\alpha}^{Ik}}+\bar{K}_{\alpha}^{Ij}\frac{\delta
}{\delta\bar{G}_{\alpha}^{Ik}}\right)\right)\;,
\end{eqnarray}
and
\begin{equation}
\Delta_{cl}^{a}=\int d^{4}x\left(  gf^{abc}\Omega_{\mu}^{b}A_{\mu}%
^{c}+gf^{abc}L^{b}c^{c}+ig\bar{Y}_{\alpha}^{i}\left(  T^{a}\right)  ^{ij}%
\psi_{\alpha}^{j}-ig\bar{\psi}_{\alpha}^{i}\left(  T^{a}\right)
^{ij}Y_{\alpha}^{j}\right)\;.\label{deltaX}
\end{equation}

\item The classical rigid invariance
\begin{equation}
R^{IJ}\Sigma=0\;,
\end{equation}
where
\begin{equation}
R^{IJ}=\int d^{4}x\left(  \bar{\lambda}^{I}\frac{\delta}{\delta\bar{\xi
}^{J}}-\lambda^{J}\frac{\delta}{\delta\xi^{I}}+\bar{H}_{\alpha}^{Ik}%
\frac{\delta}{\delta\bar{G}_{\alpha}^{Jk}}-H_{\alpha}^{Jk}\frac{\delta}{\delta
G_{\alpha}^{Ik}}+J_{\alpha}^{Jk}\frac{\delta}{\delta K_{\alpha}^{Ik}}+\bar
{J}_{\alpha}^{Ik}\frac{\delta}{\delta\bar{K}_{\alpha}^{Jk}}\right)\;.
\end{equation}

\item The equations of motion of the doublet fields:
\begin{eqnarray}
\frac{\delta\Sigma}{\delta\bar{\eta}^{I}}&=&-\partial^{2}\theta^{I}+P\theta^{I}\;,\nonumber\\
\frac{\delta\Sigma}{\delta\bar{\theta}^{I}}&=&\partial^{2}\eta^{I}-P\eta^{I}-J_{\alpha}^{Ij}\psi_{\alpha}^{j}\;,\nonumber\\
\frac{\delta\Sigma}{\delta\bar{\lambda}^{I}}+K_{\alpha}^{Ij}\frac{\delta
\Sigma}{\delta\bar{Y}_{\alpha}^{j}}&=&-\partial^{2}\xi^{I}+P\xi^{I}+G_{\alpha}^{Ij}\psi_{\alpha}^{j}\;,\nonumber\\
\frac{\delta\Sigma}{\delta\bar{\xi}^{I}}+J_{\alpha}^{Ij}\frac{\delta\Sigma}{\delta\bar{Y}_{\alpha}^{j}}&=&-\partial^{2}\lambda^{I}+P\lambda^{I}+
H_{\alpha}^{Ij}\psi_{a}^{j}\;.
\end{eqnarray}

\item The Ward identity for the  source $N$:
\begin{equation}
\frac{\delta\Sigma}{\delta N}=0\;.\label{N}
\end{equation}
\end{itemize}

\subsection{The invariant counterterm}

In order to characterize the most general invariant counterterm
which can be freely added  to all orders in perturbation theory
\cite{Piguet:1995er}, we perturb the classical action $\Sigma$ by
adding an integrated local polynomial $\Sigma^{count}$ of
dimension bounded by four, and with vanishing ghost number as well
as $Q_{4N}$ charge. We demand thus the perturbed action,
$(\Sigma+\epsilon\Sigma^{count})$, where $\epsilon$ is an
expansion parameter, fulfills, to the first order in $\epsilon$,
the same Ward identities fulfilled by the classical action
$\Sigma$, {\it \i.e.} eqs.\eqref{st}-\eqref{N}. This requirement
gives rise to the following constraints for the counterterm
$\Sigma^{count}$:
\begin{eqnarray}
{\cal B}_{\Sigma}\Sigma^{count}&=&0\;,\label{6}\\
\frac{\delta}{\delta b^{a}}\Sigma^{count}&=&0\;,\label{7}\\
\left(  \frac{\delta}{\delta\bar{c}^{a}}+\partial_{\mu}\frac{\delta}{\delta\Omega_{\mu}^{a}}\right)\Sigma^{count}&=&0\;,\label{a}\\
{\cal G}^{a}\Sigma^{count}&=&0\;,\label{b}\\
R^{IJ}\Sigma^{count}&=&0\;,\label{c}\\
\frac{\delta\Sigma^{count}}{\delta N}&=&0\;,\label{d}\\
\frac{\delta\Sigma^{count}}{\delta\bar{\eta}^{I}} &=&0\;,\\
\frac{\delta\Sigma^{count}}{\delta\bar{\theta}^{I}} &  =&0\;,\\
\frac{\delta\Sigma^{count}}{\delta\bar{\lambda}^{I}}+K_{\alpha}^{Ij}\frac{\delta\Sigma^{count}}{\delta\bar{Y}_{\alpha}^{j}} &  =&0\;,\\
\frac{\delta\Sigma^{count}}{\delta\bar{\xi}^{I}}+J_{\alpha}^{Ij}\frac{\delta\Sigma^{count}}{\delta\bar{Y}_{\alpha}^{j}} &  =&0\;.\label{12}
\end{eqnarray}
where the operator ${\cal B}_{\Sigma}$ in eq.\eqref{6} stands for
the nilpotent linearized Slavnov-Taylor operator,
\begin{eqnarray}
{\cal B}_{\Sigma}&=&\int d^{4}x\left(\frac{\delta\Sigma}{\delta\Omega_\mu^a}\frac{\delta}{\delta A_\mu^a}
+\frac{\delta\Sigma}{\delta A_\mu^a}\frac{\delta}{\delta\Omega_\mu^a}
-\frac{\delta\Sigma}{\delta\bar{\psi}_\alpha^i}\frac{\delta}{\delta Y_\alpha^i}
-\frac{\delta\Sigma}{\delta Y_\alpha^i}\frac{\delta}{\delta\bar{\psi}_\alpha^i}
+\frac{\delta\Sigma}{\delta\bar{Y}_\alpha^i}\frac{\delta}{\delta\psi_\alpha^i}
+\frac{\delta\Sigma}{\delta\psi_\alpha^i}\frac{\delta}{\delta\bar{Y}_\alpha^i}\right.\nonumber\\
&+&\left.\frac{\delta\Sigma}{\delta c^a}\frac{\delta}{\delta L^a}
+\frac{\delta\Sigma}{\delta L^a}\frac{\delta}{\delta c^a}+ib^a\frac{\delta}{\delta\bar{c}^a}
+\bar{\theta}^I\frac{\delta}{\delta\bar{\xi}^I}+ \bar{\lambda}^I\frac{\delta}{\delta\bar{\eta}^I}
+\theta^I\frac{\delta}{\delta\xi^I}+\lambda^I\frac{\delta}{\delta\eta^I}\right.\nonumber\\
&+&\left.H_\alpha^{Ij}\frac{\delta}{\delta J_\alpha^{Ij}}
+G_\alpha^{Ij}\frac{\delta}{\delta K_\alpha^{Ij}}
+\bar{H}_\alpha^{Ij}\frac{\delta}{\delta\bar{J}_\alpha^{Ij}}
+\bar{G}_\alpha^{Ij}\frac{\delta}{\delta\bar{K}_\alpha^{Ij}}
+P\frac{\delta}{\delta N}\right)\;.
\end{eqnarray}\\\\
The first constraint, eq.\eqref{6}, identifies the invariant
counterterm as the solution of the cohomology of the operator
${\cal B}_\Sigma$ in the space of the integrated local field
polynomials of dimension four. From the general results on the
cohomology of Yang-Mills theories \cite{Piguet:1995er}, it follows
that $\Sigma^{count}$ can be written as
\begin{equation}
\Sigma^{count}=\frac{a_0}{4}\int d^{4}xF_{\mu\nu}^{a}F_{\mu\nu}^{a}
+{\cal B}_{\Sigma}\Delta^{\left(  -1\right)  }\;,\label{count00}
\end{equation}
where $\Delta^{(-1)}$ is a local integrated polynomial in all
fields and sources, with dimension four, ghost number minus one
and vanishing $Q_{4N}$ charge, namely
\begin{eqnarray}
\Delta^{(-1)}&=&\int d^{4}x\left(a_{1}A_{\mu}^{a}\Omega_{\mu}^{a}+a_{2}
\partial_{\mu}\bar{c}^{a}A_{\mu}^{a}+a_{3}c^{a}L^{a}+a_{4}\bar{c}^{a}
b^{a}+a_{5}\frac{g}{2}f^{abc}c^{a}\bar{c}^{b}\bar{c}^{c}+a_{6}\bar{\lambda
}^{I}K_{\alpha}^{Ij}\psi_{\alpha}^{j}\right.\nonumber\\
&+&\left.a_{7}\bar{\psi}_{\alpha}^{j}\bar{K}_{\alpha}^{Ij}\lambda^{I}+a_{8}
\bar{\eta}^{I}G_{\alpha}^{Ij}\psi_{\alpha}^{j}+a_{9}\bar{\psi}_{\alpha}
^{j}\bar{G}_{\alpha}^{Ij}\eta^{I}+a_{10}\bar{\theta}^{I}K_{\alpha}
^{Ij}Y_{\alpha}^{j}+a_{11}\bar{Y}_{\alpha}^{j}\bar{K}_{\alpha}^{Ij}\theta
^{I}+a_{12}\bar{\xi}^{I}G_{\alpha}^{Ij}Y_{\alpha}^{j}\right.\nonumber\\
&+&\left.a_{13}\bar{Y}_{\alpha}^{j}\bar{G}_{\alpha}^{Ij}\xi^{I}+a_{14}
\bar{\xi}^{I}J_{\alpha}^{Ij}\psi_{\alpha}^{j}+a_{15}\bar{\psi}_{\alpha
}^{j}\bar{J}_{\alpha}^{Ij}\xi^{I}+a_{16}\bar{\psi}_{\alpha}^{i}Y_{\alpha
}^{i}+a_{17}\bar{Y}_{\alpha}^{i}\psi_{\alpha}^{i}+a_{18}\bar{H}_{\alpha}
^{Ij}J_{\alpha}^{Ij}+a_{19}\bar{J}_{\alpha}^{Ij}H_{\alpha}^{Ij}\right.\nonumber\\
&+&\left.a_{20}c^{a}\bar{c}^{a}K_{\alpha}^{Ij}\bar{G}_{\alpha}^{Ij}+a_{21}c^{a}
\bar{c}^{a}N~+a_{22}A_{\mu}^{a}A_{\mu}^{a}N+a_{23}A_{\mu}^{a}A_{\mu}
^{a}G_{\alpha}^{Ij}\bar{K}_{\alpha}^{Ij}+a_{24}G_{\alpha}^{Ij}\partial^{2}
\bar{K}_{\alpha}^{Ij}+a_{25}G_{\alpha}^{Ij}\bar{G}_{\alpha}^{Ij}N\right.\nonumber\\
&+&\left.a_{27}\bar{\eta}^{I}\partial^{2}\xi^{I}+a_{28}\bar{\xi}
^{I}\partial^{2}\eta^{I}+a_{29}A_{\mu}^{a}A_{\mu}^{a}\bar{\eta}^{I}\xi
^{I}+a_{30}A_{\mu}^{a}A_{\mu}^{a}\bar{\xi}^{I}\eta^{I}+a_{31}c^{a}\bar
{c}^{a}\bar{\eta}^{I}\xi^{I}\right.\nonumber\\
&+&\left.a_{32}c^{a}\bar{c}^{a}\bar{\xi}^{I}\eta^{I}+a_{33}P\bar{\eta}
^{I}\xi^{I}+a_{34}P\bar{\xi}^{I}\eta^{I}+a_{35}c^{a}\bar{c}^{a}\bar
{K}_{\alpha}^{Ij}G_{\alpha}^{Ij}+a_{36}\bar{G}_{\alpha}^{Ij}\partial
^{2}K_{\alpha}^{Ij}+a_{37}PN\right.\nonumber\\
&+&\left.a_{38}~\bar{\xi}^{I}\lambda^{I}N+a_{39}\bar{\lambda}^{I}\xi
^{I}N+a_{40}\bar{\theta}^{I}\eta^{I}N+a_{41}\bar{\eta}^{I}\theta^{I}
N+a_{42}A_{\mu}^{a}A_{\mu}^{a}\bar{G}_{\alpha}^{Ij}K_{\alpha}^{Ij}\right)\;.\label{count01}
\end{eqnarray}
 The coefficients
$\{a_0,a_1,a_2,\ldots,a_{42}\}$ in expressions \eqref{count00} and
\eqref{count01} stand for arbitrary constants parameters.\\\\
After a straightforward analysis using the conditions
\eqref{6}-\eqref{12} it turns out that  the only non-vanishing
coefficients are: $a_0$, $a_1$, $a_2$, $a_6$, $a_7$, $a_{14}$,
$a_{15}$, $a_{16}$, $a_{17}$, $a_{37}$, with the following
relations between them
\begin{equation}
a_2=a_1\;,
\end{equation}
and
\begin{eqnarray}
a_{16}&=&a_{15}\;\;=\;\;a_7\;,\nonumber\\
-a_{17}&=&a_{14}\;\;=\;\;a_6\;.
\end{eqnarray}
Then, after redefining
\begin{eqnarray}
a_6+a_7&\mapsto&a_2\;,\nonumber\\
a_{37}&\mapsto&\sigma a_3\;,
\end{eqnarray}
for the form of the final allowed counterterm one finds
\begin{eqnarray}
\Sigma^{count}&=&\int d^{4}x\left(\left(  \frac{a_0+4a_{1}}{4}\right)
F_{\mu\nu}^{a}F_{\mu\nu}^{a}-a_{1}\partial_{\mu}c^{a}\Omega_{\mu}^{a}%
-a_{1}\partial_{\mu}c^{a}\partial_{\mu}\bar{c}^{a}-a_{1}ig\bar{\psi}_{\alpha
}^{i}\left(  \gamma_{\mu}\right)  _{\alpha\beta}T^{aij}\psi_{\beta}^{j}A_{\mu
}^{a}\right.\nonumber\\
&-&\left.a_{2}\psi_{\alpha}^{i}D_{\mu}^{ij}\left(  \gamma_{\mu}\right)
_{\alpha\beta}\psi_{\beta}^{j}+a_{3}\sigma P^{2}\right)\;,\label{co}
\end{eqnarray}
which corresponds to the usual Yang-Mills counterterm in the
Landau gauge with the addition of an energy vacuum term, $\sigma
P^{2}$, related to the mass $m$.

\subsection{Stability}

It remains now to discuss the stability of the model, {\it i.e.}
to check that the counterterm $\Sigma^{count}$ can be reabsorbed
in the classical action $\Sigma$ by means of a multiplicative
redefinition of the coupling constant $g$, of the parameter
$\sigma$, anf of the fields and sources \cite{Piguet:1995er},
namely
\begin{equation}
\Sigma(g,\sigma,\phi,\Phi)+\epsilon\Sigma^{count}=\Sigma(g_{0},\sigma_{0}%
,\phi_{0},\Phi_{0})+O(\epsilon^2)\;,\label{B}
\end{equation}
where $\phi$ stands for all fields and $\Phi$ for the sources,
\begin{eqnarray}
\phi&\in&\{A,\psi,c,\bar{c},b,\xi,\bar{\xi},\lambda,\bar{\lambda},\theta,\bar{\theta},\eta,\bar{\eta}\}\;,\nonumber\\
\Phi&\in&\{\Omega,L,Y,\bar{Y},J,\bar{J},K,\bar{K},G,\bar{G},H,\bar{H},N,P\}\;.
\end{eqnarray}
Thus, by defining
\begin{eqnarray}
\phi_0&=&Z_\phi^{1/2}\phi\;,\nonumber\\
\Phi_0&=&Z_\Phi\Phi\;,\nonumber\\
g_0&=&Z_gg\;,\nonumber\\
\sigma_0&=&Z_\sigma\sigma\;,
\end{eqnarray}
we obtain
\begin{eqnarray}
Z_A^{1/2}&=&1+\frac{\epsilon}{2}\left(a_0+2a_1\right)\;,\nonumber\\
Z_g&=&1-\epsilon\frac{a_0}{2}\;,\nonumber\\
Z_{\psi}^{1/2}&=&Z_{\bar{\psi}}^{1/2}=1+\frac{\epsilon}{2}a_2\;,\nonumber\\
Z_{\sigma}&=&1+\epsilon a_3\;.\label{ren00}
\end{eqnarray}
Expressions \eqref{ren00} constitute the independent
renormalization factors. All the remaining factors can be
expressed in terms of the renormalization factors appearing in
eq.\eqref{ren00}. In fact, for the Lagrange multiplier and
Faddeev-Popov ghost fields we have
\begin{eqnarray}
Z_b&=&Z_A^{-1/2}\;,\nonumber\\
Z_c^{1/2}&=&Z_{\bar{c}}^{1/2}\;\;=\;\;Z_{g}^{-1/2}Z_{A}^{-1/4}\;,\label{ren01}
\end{eqnarray}
while the renormalization of the external BRST sources are found
\begin{eqnarray}
Z_{\Omega}&=&Z_{g}^{-1/2}Z_{A}^{-1/4}\;,\nonumber\\
Z_L&=&Z_A^{1/2}\;,\nonumber\\
Z_{\bar{Y}}&=&Z_Y\;\;=\;\;Z_g^{-1/2}Z_{A}^{1/4}Z_{\psi}^{-1/2}\;.\label{ren02}
\end{eqnarray}
As expected from \eqref{co}, the renormalization properties of the usual Yang-Mills sector are preserved.
For the doublet fields we obtain
\begin{eqnarray}
Z_{\xi}^{1/2}&=&Z_{\xi}^{1/2}\;\;=\;\;Z_{\lambda}^{1/2}\;\;=\;\;Z_{\lambda}^{1/2}\;\;=\;\;1\;,\nonumber\\
Z_{\theta}^{1/2}&=&Z_{\bar{\theta}}^{1/2}\;\;=\;\;Z_{\eta}^{-1/2}\;\;=\;\;Z_{\bar{\eta}}^{-1/2}\;\;=\;\;Z_{g}^{1/2}Z_{A}^{-1/4}\;.\label{ren03}
\end{eqnarray}
Finally, for the remaining sources we have
\begin{eqnarray}
Z_H&=&Z_{\bar{H}}\;\;=\;\;Z_{G}\;\;=\;\;Z_{\bar{G}}\;\;=\;\;Z_{\psi}^{-1/2}\;,\nonumber\\
Z_{J}&=&Z_{\bar{J}}\;\;=\;\;Z_{K}\;\;=\;\;Z_{\bar{K}}\;\;=\;\;Z_{g}^{-1/2}Z_{A}^{1/4}Z_{\psi}^{-1/2}\;,\nonumber\\
Z_P&=&1\;.\label{ren04}
\end{eqnarray}\\\\
This ends the proof of the multiplicative renormalizability of the
model proposed in this article. For completeness, let us give the
expression of the bare action written in terms of the renormalized
fields and parameters:
\begin{eqnarray}
\Sigma&=&\int d^{4}x\biggl( \frac{1}{2}Z_A\left( \partial _{\mu
}A_{\nu }^{a}-\partial _{\nu }A_{\mu }^{a}\right) \partial _{\mu
}A_{\nu
}^{a}+Z_gZ_{A}^{3/2}gf^{amn}\partial _{\mu }A_{\nu }^{a}A_{\mu }^{m}A_{\nu }^{n}+%
\frac{1}{4}Z_{g}^{2}Z_{A}^{2}g^{2}f^{abc}f^{amn}A_{\mu }^{b}A_{\nu }^{c}A_{\mu
}^{m}A_{\nu }^{n}\nonumber\\
&+&Z_\psi\bar{\psi}_\alpha^i\left(\gamma_\mu\right)_{\alpha\beta}\partial_\mu\psi_\beta^i
-iZ_gZ_{\psi}Z_{A}^{1/2}g\bar{\psi}_{\alpha }^{i}\left( \gamma_{\mu
}\right)_{\alpha \beta}A_{\mu }^{a}\left(
T^{a}\right)^{ij}\psi_{\beta}^{j} + ib^{a}\partial_{\mu }A_{\mu
}^{a} + ig\bar{Y}_{\alpha }^{i}c^{a}\left(
T^{a}\right)^{ij}\psi_{\alpha }^{j}\nonumber\\
&-&ig\bar{\psi}_{\alpha }^{j}c^{a}\left( T^{a}\right)^{ji}Y_{\alpha
}^{i} -Z_{g}^{-1}Z_{A}^{-1/2}\left( \partial_{\mu
}\bar{c}^{a}+\Omega _{\mu }^{a}\right) \partial _{\mu
}c^{a}+\left(\partial _{\mu}\bar{c}^{a}+\Omega _{\mu }^{a}\right)
gf^{abc}A_{\mu }^{c}c^{b}+\frac{1}{2}%
gL^{a}f^{abc}c^{b}c^{c} \nonumber\\
&-&\bar{\lambda}^{I}\partial ^{2}\xi^{I}-%
\bar{\xi}^{I}\partial ^{2}\lambda^{I}-\bar{\eta}^{I}\partial^{2}\theta^{I}
+\bar{\theta}^{I}\partial^{2}\eta^{I}+m\bar{\lambda}^{I}\xi^{I}+m\bar{\xi}^{I}\lambda^{I}
+m\bar{\eta}^{I}\theta^{I}-m\bar{%
\theta}^{I}\eta^{I} \nonumber\\
&+& M_{1}^{2}\bar{\xi}^{i}_{\alpha}\psi_{\alpha
}^{i}+M_{1}^{2}\bar{\psi}_{\alpha }^{i}\xi _{\alpha
}^{i}-M_2\bar{\lambda}_{\alpha }^{i}\psi _{\alpha
}^{i}-M_2\bar{\psi}_{\alpha }^{i}\lambda _{\alpha
}^{i}+Z_\sigma\sigma m^{2}\biggr)\;.\label{renorm}
\end{eqnarray}
From expression \eqref{renorm} one can obtain the renormalized
version of \eqref{integrated0}, namely
\begin{equation}
S_{\psi}^{\mathrm{ren}}=\int d^{4}x\left(Z_\psi\bar{\psi}_{\alpha}^{i}\left(  \gamma_{\mu}\right)  _{\alpha\beta}\partial_{\mu
}\psi_{\beta}^{i}-iZ_gZ_\psi Z_A^{1/2} g\bar{\psi}^i_\alpha(\gamma_\mu)_{\alpha\beta}A_\mu^a(T^a)^{ij}\psi^j_\beta-2Z_\psi M_1^2 M_2\bar{\psi}_\alpha^i\left( \frac{1}{\partial^{2}-m^{2}}\right) \psi_{\alpha}^{i}\right)\;.\label{integrated1}
\end{equation}

\section{Inclusion of the Gribov-Zwanziger term}
\label{GZ_term}

\subsection{A brief overview of the Gribov-Zwanziger action and of its soft BRST breaking term}
The Gribov-Zwanziger framework
\cite{Gribov:1977wm,Zwanziger:1989mf,Zwanziger:1992qr} enables one
to take into account the existence of the Gribov copies, which
affect the Landau gauge\footnote{See \cite{Sobreiro:2005ec} for an
introduction to the subject of the Gribov copies.}. This is done
by restricting the domain of integration in the Feynman path
integral to the so called Gribov region $\Omega$, defined as the
set of fields fulfilling the Landau gauge condition and for which
the Faddeev-Popov operator,
${\cal{M}}^{ab}=-\partial_{\mu}D^{ab}_{\mu}(A)$, is strictly
positive
\begin{equation}
\Omega=\{ A^a_{\mu} \;\; , \;\;\partial_\mu A^a_\mu=0 \;\; .
\;\;{\cal{M}}^{ab}
> 0\} \; , \label{om}
\end{equation}
As shown in \cite{Zwanziger:1989mf,Zwanziger:1992qr}, the
implementation of the restriction to the region $\Omega$ is done
by adding to the starting action a nonlocal term, known as the
horizon function, namely
\begin{equation}
S_{\mathrm{GZ}}=-g^{2}\gamma^{4}\int{d^{4}x}\,f^{abc}A^{b}_{\mu}\left[(\partial\cdot
D)^{-1}\right]^{ad}f^{dec}A^{e}_{\mu}\;. \label{hz}
\end{equation}
The parameter $\gamma$ has the dimension of a mass and is known as
the Gribov parameter. It is not a free parameter, being determined
in a self consistent way through the gap equation
\begin{equation}
\frac{\delta \Gamma}{\delta \gamma^2}= 0 \; , \label{gap}
\end{equation}
where $\Gamma$ stands for the effective action evaluated in the
presence of the horizon function (\ref{hz}). Despite of its
nonlocal character, the term $S_{\mathrm{GZ}}$ can be cast in
local form by introducing a suitable set of auxiliary fields
$(\varphi^{ab}_{\mu},\bar\varphi^{ab}_{\mu},\omega^{ab}_{\mu},\bar\omega^{ab}_{\mu})$,
\begin{eqnarray}
e^{-S_{\mathrm{GZ}}}&=&\int D\varphi D\bar\varphi\,D\omega
D\bar\omega\;e^{-S_{\mathrm{GZ}}^{\mathrm{Local}}}\;,\\
S_{\mathrm{GZ}}^{\mathrm{Local}}&=&\int{d^{4}x}\,\left(-\bar\varphi^{ac}_{\mu}\partial_{\nu}D^{ab}_{\nu}\varphi^{bc}_{\mu}
+\bar\omega^{ac}_{\mu}\partial_{\nu}D^{ab}_{\nu}\omega^{bc}_{\mu}+(\partial_{\nu}\bar\omega^{ac}_{\mu})
gf^{abd}\varphi^{bc}_{\mu}D^{de}_{\nu}c^{e}\right)\nonumber\\
&+&g\gamma^{2}\int{d^{4}x}\,\biggl(f^{abc}(\varphi^{ab}_{\mu}-\bar\varphi^{ab}_{\mu})A^{c}_{\mu}
-\frac{4}{g}(N^{2}-1)\gamma^{2}\biggr)\;.\label{GZ_local}
\end{eqnarray}
Here, $(\varphi^{ab}_{\mu},\bar\varphi^{ab}_{\mu})$ form a pair of
complex commuting fields, while
$(\omega^{ab}_{\mu},\bar\omega^{ab}_{\mu})$ form a pair of complex
anti-commuting fields. These fields are assembled in BRST doublets
\begin{eqnarray}
&s\varphi^{ab}_{\mu}=\omega^{ab}_{\mu}\,,\qquad
s\omega^{ab}_{\mu}=0\,,&\nonumber\\
&s\bar\omega^{ab}_{\mu}=\bar\varphi^{ab}_{\mu}\;,\qquad
s\bar\varphi^{ab}_{\mu}=0\,,&\label{brs2}
\end{eqnarray}
and, as pointed out in
\cite{Dudal:2005na,Dudal:2007cw,Dudal:2008sp,Baulieu:2008fy}, the
local action (\ref{GZ_local}) gives rise to a soft breaking of the
BRST symmetry, due to the presence of the Gribov parameter
$\gamma$. In fact, it turns out that expression ({\ref{GZ_local}})
can be written as
\begin{equation}
S_{\mathrm{GZ}}^{\mathrm{Local}}=S_{\varphi\omega} +
g\gamma^{2}\int{d^{4}x}\,\biggl(f^{abc}(\varphi^{ab}_{\mu}-\bar\varphi^{ab}_{\mu})A^{c}_{\mu}
-\frac{4}{g}(N^{2}-1)\gamma^{2}\biggr) \;, \label{q1}
\end{equation}
with
\begin{eqnarray}
S_{\varphi\omega}&=&-s\int{d^{4}x}\,\bar\omega^{ac}_{\mu}\partial_{\nu}D^{ab}_{\nu}\varphi^{bc}_{\mu}\nonumber\\
&=&\int{d^{4}x}\,\left(-\bar\varphi^{ac}_{\mu}\partial_{\nu}D^{ab}_{\nu}\varphi^{bc}_{\mu}
+\bar\omega^{ac}_{\mu}\partial_{\nu}D^{ab}_{\nu}\omega^{bc}_{\mu}+(\partial_{\nu}\bar\omega^{ac}_{\mu})
gf^{abd}\varphi^{bc}_{\mu}D^{de}_{\nu}c^{e}\right)\;,
\end{eqnarray}
so that
\begin{eqnarray}
sS_{\mathrm{GZ}}^{\mathrm{Local}}&=&\gamma^{2}\Delta_{\gamma}\;,\nonumber\\
\Delta_{\gamma}&=&\int{d^{4}x}\,\Bigl(gf^{abc}\omega^{ab}_{\mu}A^{c}_{\mu}
-gf^{abc}(\varphi^{ab}_{\mu}-\bar\varphi^{ab}_{\mu})D^{cd}_{\mu}c^{d}\Bigr)\;.
\end{eqnarray}
In order to keep control of the soft BRST breaking term, we
proceed as before and introduce a set of external sources
$(U^{ab}_{\mu\nu},\bar{U}^{ab}_{\mu\nu},V^{ab}_{\mu\nu},\bar{V}^{ab}_{\mu\nu})$
transforming as
\begin{eqnarray}
&sV^{ab}_{\mu\nu}=U^{ab}_{\mu\nu}\,,\qquad
sU^{ab}_{\mu\nu}=0\,,&\nonumber\\
&s\bar{U}^{ab}_{\mu\nu}=\bar{V}^{ab}_{\mu\nu}\,,\qquad
s\bar{V}^{ab}_{\mu\nu}=0\,,&\label{brs3}
\end{eqnarray}
and whose physical values are defined by
\begin{eqnarray}
&V^{ab}_{\mu\nu}\Bigl|_{\mathrm{phys}}=-\bar{V}^{ab}_{\mu\nu}\Bigl|_{\mathrm{phys}}
=-\gamma^{2}\delta^{ab}\delta_{\mu\nu}\,,&\nonumber\\
&U^{ab}_{\mu\nu}\Bigl|_{\mathrm{phys}}=\bar{U}^{ab}_{\mu\nu}\Bigl|_{\mathrm{phys}}
=0\,.&\label{phys_val_UV}
\end{eqnarray}
Thus, we can replace the breaking term in eq.\eqref{q1} by the
following BRST-invariant source term
\begin{eqnarray}
S_{\mathrm{source}}&=&s\int{d^{4}x}\,\Bigl(\bar{U}^{ab}_{\mu\nu}D^{ac}_{\mu}{\varphi}^{cb}_{\nu}
+V^{ab}_{\mu\nu}D^{ac}_{\mu}\bar{\omega}^{cb}_{\nu}
-\bar{U}^{ab}_{\mu\nu}V^{ab}_{\mu\nu}\Bigr)\nonumber\\
&=&\int{d^{4}x}\,\Bigl(\bar{V}^{ab}_{\mu\nu}D^{ac}_{\mu}{\varphi}^{cb}_{\nu}
-\bar{U}^{ab}_{\mu\nu}[D^{ac}_{\mu}\omega^{cb}_{\nu}
+gf^{acd}(D^{de}_{\mu}c^{e})\varphi^{cb}_{\nu}]\nonumber\\
&+&U^{ab}_{\mu\nu}D^{ac}_{\mu}\bar{\omega}^{cb}_{\nu}
+V^{ab}_{\mu\nu}[D^{ac}_{\mu}\bar\varphi^{cb}_{\nu}
+gf^{acd}(D^{de}_{\mu}c^{e})\bar\omega^{cb}_{\nu}]
-(\bar{U}^{ab}_{\mu\nu}U^{ab}_{\mu\nu}
-\bar{V}^{ab}_{\mu\nu}V^{ab}_{\mu\nu})\Bigl)\;.
\end{eqnarray}
Notice that the source term $S_{\mathrm{source}}$ gives back the
original BRST soft breaking term when the sources attain their
physical values \eqref{phys_val_UV}. In fact, after a little
algebra, one finds
\begin{equation}
S_{\mathrm{source}}\Bigl|_{\mathrm{phys}}=g\gamma^{2}\int{d^{4}x}\,\biggl(f^{abc}(\varphi^{ab}_{\mu}
-\bar\varphi^{ab}_{\mu})A^{c}_{\mu}
-\frac{4}{g}(N^{2}-1)\gamma^{2}\biggr)\;.
\end{equation}
We are now ready  to discuss the inclusion of the Gribov-Zwanziger
term \eqref{GZ_local} into our starting action $\Sigma$,
eq.\eqref{sigma}. To that purpose we consider the more general
action
\begin{equation}
\Sigma_{\mathrm{tot}}=\Sigma+S_{\varphi\omega}+S_{\mu}+S_{\mathrm{source}}\;,\label{actionX}
\end{equation}
where
\begin{equation}
S_{\mu}=\mu^{2}\,s\int{d^{4}x}\,\bar\omega^{ab}_{\mu}\varphi^{ab}_{\mu}
=\mu^{2}\int{d^{4}x}\,\Bigl(\bar\varphi^{ab}_{\mu}\varphi^{ab}_{\mu}
-\bar\omega^{ab}_{\mu}\omega^{ab}_{\mu}\Bigr)\;.
\end{equation}
As discussed in \cite{Dudal:2007cw,Dudal:2008sp}, the term
$S_{\mu}$ takes into account the nontrivial dynamics of the
auxiliary localizing fields
$(\bar\varphi^{ab}_{\mu}\varphi^{ab}_{\mu})$.  The introduction of
the BRST-invariant term $S_{\mu}$ follows from the observation
that the dimension two condensate
$\langle\bar\varphi^{ab}_{\mu}\varphi^{ab}_{\mu}
-\bar\omega^{ab}_{\mu}\omega^{ab}_{\mu}\rangle$ has a nonzero
value for non-vanishing Gribov parameter $\gamma$, namely
\begin{equation}
\langle\bar\varphi^{ab}_{\mu}\varphi^{ab}_{\mu}
-\bar\omega^{ab}_{\mu}\omega^{ab}_{\mu}\rangle=\frac{3(N^{2}-1)}{64\pi}2^{1/2}gN^{1/2}\gamma^2\;.
\end{equation}
The existence of this condensate is taken into account through the
mass parameter $\mu$ which, in a way similar to the Gribov
parameter $\gamma$, is determined by a variational principle, see
\cite{Dudal:2007cw,Dudal:2008sp}.

\subsection{Renormalizability of the quark-gluon model in the presence of the Gribov-Zwanziger term}

In order to discuss the renormalizability of expression
\eqref{actionX},  one uses
the  Ward identities that  have already been established in
\cite{Zwanziger:1992qr,Maggiore:1993wq,Dudal:2005na,Dudal:2007cw,Dudal:2008sp}.
Moreover,
it exhibits the following symmetry
\begin{equation}
\Theta_{\mu\nu}^{ab}\Sigma_{\mathrm{tot}}=0 \;, \label{nsymm}
\end{equation}
where the operator $\Theta_{\mu\nu}^{ab}$ is given by
\begin{eqnarray}
\Theta_{\mu\nu}^{ab}&=&\int d^{4}x\left(\varphi_\mu^{ac}\frac{\delta}{\delta\varphi_{\nu}^{bc}}-\bar{\varphi}_\nu^{bc}\frac{\delta}{\delta\bar{\varphi}_\mu^{ac}}+ \omega_\mu^{ac}\frac{\delta}{\delta\omega_{\nu}^{bc}}-\bar{\omega}_\nu^{bc}\frac{\delta}{\delta\bar{\omega}_\mu^{ac}}+ V_{\mu\sigma}^{ac}\frac{\delta}{\delta V_{\nu\sigma}^{bc}}-\bar{V}_{\mu\sigma}^{ac}\frac{\delta}{\delta\bar{V}_{\nu\sigma}^{bc}}\right.
\nonumber\\
&+&\left.U_{\mu\sigma}^{ac}\frac{\delta}{\delta U_{\nu\sigma}^{bc}}-\bar{U}_{\mu\sigma}^{ac}\frac{\delta}{\delta\bar{U}_{\nu\sigma}^{bc}}\right).\label{qop2}
\end{eqnarray}
The Ward identity \eqref{nsymm} expresses the invariance of the
action \eqref{actionX} under a global $U(4(N^2-1))$
transformation. This symmetry works exactly as the global $U(4N)$
symmetry associated with the spinor sector of the theory through
the operator \eqref{qop}. These global symmetries ensure in fact
that no mixing terms between the two set of BRST doublet fields,
{\it i.e.}
$(\bar\varphi^{ab}_{\mu},\varphi^{ab}_{\mu},\bar\omega^{ab}_{\mu},\omega^{ab}_{\mu})$
and $(\xi_\alpha^i, \theta_\alpha^i, \eta_\alpha^i,
\lambda_\alpha^i)$, arise in the allowed counterterm. All Ward
identities of the Gribov-Zwanziger action remain valid in the
present case. This is also the case of the identities
\eqref{st}-\eqref{N}. Of course, the Slavnov-Taylor identity \eqref{st} is supplemented by suitable extra terms accounting for the BRST new doublets of the gluon sector, \eqref{brs2} and \eqref{brs3},
\begin{equation}
{\cal S}(\Sigma)\rightarrow {\cal S}(\Sigma_{\mathrm{tot}})+\int{d^4x}\left(\omega_\mu^{ab}\frac{\delta\Sigma_{\mathrm{tot}}}{\delta \varphi_\mu^{ab}}+\bar{\varphi}_\mu^{ab}\frac{\delta\Sigma_{\mathrm{tot}}}{\delta\bar{\omega}_\mu^{ab}}+ U_{\mu\nu}^{ab}\frac{\delta\Sigma_{\mathrm{tot}}}{\delta V_{\mu\nu}^{ab}}+ \bar{V}_{\mu\nu}^{ab}\frac{\delta\Sigma_{\mathrm{tot}}}{\delta \bar{U}_{\mu\nu}^{ab}}\right)\;.
\end{equation}
Also, the ghost equation \eqref{ghost} needs a little modification which, due to the presence of the Gribov-Zwanziger term, generalizes to
\begin{equation}
{\cal G}^a\rightarrow {\cal G}^a+gf^{abc}\int{d^4x}\left(\varphi_\mu^{bd}\frac{\delta}{\delta \omega_\mu^{cd}}+ \bar{\omega}_\mu^{bd}\frac{\delta}{\delta\bar{\varphi}_\mu^{cd}}+ V_{\mu\nu}^{db}\frac{\delta}{\delta U_{\mu\nu}^{dc}}+ \bar{U}_{\mu\nu}^{db}\frac{\delta}{\delta \bar{V}_{\mu\nu}^{dc}}\right)\;,
\end{equation}
while the classical breaking term \eqref{deltaX} remains
unmodified. The previous algebraic analysis can be now repeated
for the more general action \eqref{actionX}. The final output is
that the action \eqref{actionX} remains renormalizable to all
orders.

\section{Conclusion}\label{conclusion}

In this work we have considered a model that accounts for a
modification of the infrared behavior of   quark  and gluon
propagators in Yang-Mills theories. This is achieved  through the
introduction of suitable mass parameters which give rise to a soft
breaking of the BRST symmetry, as outlined in
\cite{Baulieu:2008fy}.\\\\Being soft, the breaking term can be
neglected in the   ultraviolet region, where the standard massless
quark propagator is recovered as well as the notion of exact BRST
invariance. Moreover, in the infrared region the quark propagator
turns out to be deeply modified, as shown by expression \eqref{e}.
The physical reasoning behind the introduction of the soft BRST
breaking and of the ensuing modification of the propagator relies
on quark confinement and on the breaking of the chiral symmetry,
both occurring in the non-perturbative infrared region. It is
worth remarking that the quark propagator \eqref{e} is in fact in
qualitative agreement with the fitting formulas employed in the
numerical studies of the quark two-point function through lattice
simulations in the Landau gauge
\cite{Parappilly:2005ei,Furui:2006ks}. \\\\The main result of the
present article is the analysis of the renormalizability of the
model, which we have  shown to hold at any given finite   orders
of perturbation theory, by making use of the algebraic
renormalization \cite{Piguet:1995er}. The inclusion of the
Gribov-Zwanziger term which enables us to implement the
restriction to the Gribov region $\Omega$ has also been taken into
account. Despite the presence of the soft BRST breaking term, the
renormalizability of the model is guaranteed by the large set of
Ward identities, eqs.\eqref{st}-\eqref{N}, which can be
established.
\\\\
We expect that the mechanism of introducing non-perturbative
infrared effects through the soft breaking of the BRST symmetry
\cite{Baulieu:2008fy} applies as well to other kinds of models,
including supersymmetric and topological field theories.

\section*{Acknowledgments}
The Conselho Nacional de Desenvolvimento Cient\'{\i}fico e Tecnol\'{o}gico
(CNPq-Brazil), the Faperj, Funda{\c{c}}{\~{a}}o de Amparo {\`{a}} Pesquisa do
Estado do Rio de Janeiro, the Latin American Center for Physics (CLAF) the
SR2-UERJ and the Coordena{\c{c}}{\~{a}}o de Aperfei{\c{c}}oamento de Pessoal
de N{\'{\i}}vel Superior (CAPES) are gratefully acknowledged for financial support.
This work has been partially supported by the contract ANR (CNRS-USAR), \texttt{05-BLAN-0079-01}.

\end{document}